\documentclass[10pt,preprint,twocolumn]{aastex62}
\newcommand{\roma}[1]{\uppercase\expandafter{\romannumeral#1}}

\usepackage{amsmath}
\usepackage{subfigure}
\usepackage{hyperref}

\usepackage{lineno}
\usepackage{color}
\usepackage{makecell}
\usepackage{graphicx}
\usepackage{natbib}
\usepackage{amssymb,txfonts}
\usepackage{multirow}
\usepackage{array}

\usepackage{sidecap}
\usepackage{hyperref}
\usepackage{epstopdf}
\usepackage{footnote}
\usepackage{tabularx}
\usepackage{booktabs}

\shorttitle{Photospheric Motion Driven Two-Stage Reconnection in a Small-scale Chromospheric Jet}
\shortauthors{Tang et al.}
\submitjournal{APJL}

\begin{document}
\title{From Photospheric Footpoint Motion to Plasmoid Ejection: A Two-Stage Reconnection Process in a Small-scale Chromospheric Jet}
\correspondingauthor{Yuandeng Shen}
\email{ydshen@hit.edu.cn}

\affiliation{Yunnan Observatories, Chinese Academy of Sciences, Kunming 650216, China}
\author{Zehao Tang$^{1,4}$}
\noaffiliation

\author[0000-0001-9493-4418]{Yuandeng Shen}
\affiliation{State Key Laboratory of Solar Activity and Space Weather, School of Aerospace, Harbin Institute of Technology, Shenzhen 518055, China}
\affiliation{Shenzhen Key Laboratory of Numerical Prediction for Space Storm, Harbin Institute of Technology, Shenzhen 518055, China}



\author{Chengrui Zhou}
\affiliation{Yunnan Observatories, Chinese Academy of Sciences, Kunming 650216, China}
\affiliation{University of Chinese Academy of Sciences, Beijing, 100049, China}

\author{Surui Yao}
\affiliation{Yunnan Observatories, Chinese Academy of Sciences, Kunming 650216, China}
\affiliation{University of Chinese Academy of Sciences, Beijing, 100049, China}

\author{Dongxu Liu}
\affiliation{Yunnan Observatories, Chinese Academy of Sciences, Kunming 650216, China}
\affiliation{University of Chinese Academy of Sciences, Beijing, 100049, China}

\author{Xiaobo Li}
\affiliation{Yunnan Observatories, Chinese Academy of Sciences, Kunming 650216, China}

\begin{abstract}
Using high spatiotemporal resolution, multi-wavelength observations from the New Vacuum Solar Telescope (NVST) and the Solar Dynamics Observatory (SDO), we present a detailed analysis of a small-scale chromospheric jet driven by plasmoid-mediated magnetic reconnection. Our results reveal that the entire process is governed by the dynamic evolution of photospheric magnetic footpoints, which proceeds in two distinct stages. An initial separating motion of the footpoints corresponds to a mild reconnection phase, characterized by a short current sheet and the eruption of a cool H$\alpha$ jet. Subsequently, a converging motion of the footpoints triggers an intense reconnection phase. During this intense stage, the current sheet rapidly elongates, and the resulting decrease in its aspect ratio initiates a tearing-mode instability, forming a plasmoid. The appearance of this plasmoid mediates the onset of fast magnetic reconnection, which produces a hot EUV jet and is concurrent with significant magnetic flux cancellation. We interpret this cancellation as the submergence of newly formed, post-reconnection loops. Furthermore, we identify a distinct, high-temperature plasma blob in the jet spire, significantly hotter than the surrounding jet plasma. We attribute this feature to a secondary heating process, likely caused by reconnection between the upward-propagating plasmoid and the overlying magnetic cusp structure. These observations provide a comprehensive, observationally driven picture (from the initial photospheric triggers to the multi-stage, plasmoid-mediated reconnection) that forms chromospheric jets, highlighting the critical role of footpoint motions in solar atmospheric dynamics.
 
\end{abstract}

\section{Introduction}
The continuous and often random motion of magnetic field footpoints in the photosphere profoundly influences atmospheric magnetic activity by constantly altering large-scale magnetic connectivity. The consequences of this motion depend on the timescale of the footpoint driving relative to the timescale of current dissipation in the atmosphere, leading to two general scenarios. In the first scenario, when footpoint motions are rapid—occurring on timescales shorter than the dissipation time—the magnetic field cannot relax accordingly. This quasi-statically stresses the field, gradually converting mechanical energy into free magnetic energy that is stored in structures such as coronal sheared arcades, magnetic flux ropes (MFRs) \citep{2020RAA....20..165L, 2011RAA....11..594S, 2012ApJ...750...12S, 2024ApJ...975L...5Y, 2021ApJ...921L..33Y,2007ApJ...667L.105J}, and magnetic braids as well \citep{1983ApJ...264..642P, 2013Natur.493..501C, 2018ApJ...854...80H, 2021NatAs...5...54A, 2021NatAs...5..237B, 2022A&A...667A.166C, 2023A&A...679A...9B}. In the second scenario, when footpoint motions are slow, occurring on timescales longer than the dissipation time, the magnetic field is continuously reconfigured by magnetic reconnection at tangled nodes, frequently observed as dynamic atmospheric events \citep[e.g.,][]{2025ApJ...987L...5L}, including nanoflares \citep{1988ApJ...330..474P, 2018ApJ...862L..24P, 2019ApJ...872...32S, 2025ApJ...985L..11T}, ultraviolet bursts \citep{2014Sci...346C.315P, 2014Sci...346A.315T}, and Ellerman bombs \citep{2016ApJ...824...96T, 2016ApJ...823..110R}.

The motion of photospheric footpoints also plays a direct role in the development of dissipative current sheets, the sites where magnetic reconnection occurs. Theoretical models of cancellation nanoflares suggest that the converging motion of opposite-polarity footpoints is an effective mechanism for causing coronal magnetic null points to collapse into current sheets \citep{2018ApJ...862L..24P}, a scenario recently confirmed by high-resolution observations \citep{2025ApJ...985L..11T}. Furthermore, \citet{2024MNRAS.534.3133P} demonstrated that the separation distance between footpoints is inversely correlated with the length of the dissipation region, implying that photospheric convergence contributes directly to its extension. These studies collectively establish that footpoint dynamics are a significant driver in shaping reconnection regions in the solar atmosphere.

Plasmoids, also known as magnetic islands, are critical products of these dissipative current sheets and are essential for facilitating fast magnetic reconnection. They are generated via the tearing-mode instability when the current sheet's aspect ratio (width-to-length) drops below a critical threshold \citep[estimated as $\sim$0.1 by][]{2003SoPh..217..187V}. Plasmoids have been ubiquitously observed in diverse environments, from laboratory plasmas \citep{2017NucFu..57e6037T} and the Earth's magnetotail \citep{1994GeoRL..21.2935H, 2010PhRvL.104q5003W} to the solar corona \citep{2010ApJ...713.1292M, 2012ApJ...745L...6T, 2013A&A...557A.115K, 2016NatPh..12..847L, 2017ApJ...851...67S, 2024A&A...687A.190H} and chromosphere \citep{2012ApJ...759...33S, 2017ApJ...851L...6R, 2021A&A...647A.188D, 2023A&A...673A..11R, 2024ApJ...966L..29C, 2024A&A...683A.190R, 2025A&A...696A...3L}. By mediating the transport of mass, momentum, and magnetic flux, plasmoids fundamentally alter a system's evolution. More importantly, the tearing instability that forms them provides a compelling mechanism to bridge magnetohydrodynamic (MHD) and kinetic-scale dynamics in the reconnection process \citep{2022NatRP...4..263J}.

Drifting pulsating structures (DPSs) observed in the radio spectrum are thought to be generated by Langmuir waves, which are created by electrons trapped within plasmoids. Consequently, DPSs serve as crucial indicators of tearing instability in plasma \citep{2000A&A...360..715K}. For instance, a chain of plasmoids was identified through observations of a series of DPSs \citep{2004A&A...417..325K}. Beyond radio signatures, plasmoids have been directly detected across various remote-sensing wavelengths, significantly advancing our understanding of their dynamics \citep{2016NatPh..12..847L, 2010ApJ...713.1292M, 2012ApJ...745L...6T, 2013A&A...557A.115K, 2024ApJ...966L..29C, 2024A&A...683A.190R}. Observational studies reveal that plasmoids can propagate in different directions. While early observations frequently reported anti-sunward motion \citep{2003ApJ...594.1068K, 2012PhRvX...2b1015S}, more recent studies have also identified sunward-propagating plasmoids \citep{2010ApJ...713.1292M, 2020A&A...644A.158P, 2013ApJ...767..168L}. Furthermore, \citet{2013A&A...557A.115K} even observed bidirectional plasmoid movement during a single flare event. The direction of plasmoid propagation is not random. It is primarily determined by the net tension from newly reconnected magnetic field lines at either end of the plasmoid \citep{2008A&A...477..649B}. Regarding their speed, observations using LASCO images showed plasmoids moving at 300 to 650 km s$^{-1}$, which is consistent with the local Alfvén speed \citep{2003ApJ...594.1068K}. This aligns with two-dimensional resistive MHD simulations, which indicate that upward-propagating plasmoids can achieve local Alfvén speeds, whereas downward-propagating plasmoids move at only a fraction of this velocity \citep{2008A&A...477..649B, 2018ApJ...858...70F}.

Theoretical studies propose the existence of scaling laws that govern plasmoid behavior. A key insight is the identification of their fractal properties—a spatial self-similarity that helps reduce the scale of dissipative current sheets and thus accelerates magnetic reconnection \citep{2001EP&S...53..473S}. This self-similarity results in power-law distributions for the time variations of reconnection and energy release rates. Furthermore, a scaling law for the magnetic flux of plasmoids has been proposed, though competing theories exist. One model suggests that the distribution function $f(\phi)$  of plasmoid magnetic flux $\phi$ follows an inverse-square law, $f(\phi) \sim 1/\phi^2$  \citep{2010PhRvL.105w5002U}. In contrast, other studies propose a distribution of $f(\phi) \sim 1/\phi$ in current sheets with high Lundquist numbers \citep{2012PhRvL.109z5002H}, while kinetic models predict a distribution with an exponentially decaying tail \citep{2012PhRvL.108y5005F}. Another study reported that the number of plasmoids first increases and then decreases with the widening of the current sheet \citep{2013ApJ...771L..14G}. The study of plasmoid scaling laws is significant because it offers a compelling mechanism to bridge magnetohydrodynamic (MHD) and kinetic-scale dynamics within the reconnection process.

Plasmoids are crucial in triggering powerful energy release episodes, primarily through a process of merging where smaller plasmoids coalesce into larger ones \citep{1979PhFl...22.2140P}. This merging process is facilitated by secondary current sheets that form between the plasmoids, which fragment their O-shaped magnetic field lines and contribute to additional energy release and localized heating \citep{2012A&A...541A..86K}. Beyond plasma heating, plasmoids are also central to particle acceleration. One of the most significant mechanisms is the repeated reflection of electrons between contracting plasmoids, analogous to a ball gaining energy between two converging walls \citep{2006Natur.443..553D, 2010ApJ...714..915O}. Furthermore, the merging of plasmoids with post-reconnected loops can drive further acceleration, explaining observations of loop-top X-ray sources \citep{2008A&A...477..649B, 2011ApJ...733..107K}, a scenario supported by observational evidence \citep{2010ApJ...713.1292M, 2013ApJ...767..168L}. Finally, the ejection of large plasmoids, formed from this continual coalescence, can enhance the overall magnetic reconnection rate and intensify the energy release \citep{2001EP&S...53..473S, 2009PhPl...16k2102B}, providing a promising mechanism for producing hot explosions even in the cool chromosphere \citep{2021A&A...646A..88N}.

This study utilizes high-resolution data from the NVST and SDO/AIA to investigate a small-scale, plasmoid-mediated reconnection event in association with a chromospheric jet. Our results reveal the complete evolution of this event, including the convergence of photospheric footpoints, the formation and elongation of a current sheet, the subsequent appearance of plasmoids, and an associated enhancement in the energy release rate. This study suggests that the convergence of the current sheet's photospheric footpoints facilitates its elongation, and the decreasing width-to-length ratio of the sheet drives the formation of plasmoids. Section 2 describes the observational data and methodology, Section 3 presents the detailed observational results, and Section 4 provides our conclusions and a discussion.

\section{Instruments and Data}
The reported event was simultaneously recorded by the ground-based New Vacuum Solar Telescope (NVST) and the space-based Solar Dynamics Observatory (SDO). The NVST provided high-resolution chromospheric observations at the H$\alpha$ line center (6562.8~\AA) using a Lyot filter with a 0.25~\AA\ bandpass \citep{2014RAA....14..705L}. The raw (Level 0) data were processed to Level 1 by applying dark-current and flat-field corrections. These Level 1 data were then reconstructed using a speckle masking method to produce the final Level 1+ data product \citep{2016NewA...49....8X}. For the observation period, the NVST H$\alpha$ data have an image cadence of 12~s and a pixel size of 0$\arcsec$.14. In addition, we utilized extreme ultraviolet (EUV) observations from the Atmospheric Imaging Assembly \citep[AIA;][]{2012SoPh..275...17L} and photospheric line-of-sight (LOS) magnetograms from the Helioseismic and Magnetic Imager \citep[HMI;][]{2012SoPh..275..229S}, both onboard the SDO. The SDO/AIA images have a cadence and pixel size of 12 s and 0\arcsec.6, respectively, while the SDO/HMI magnetograms have a cadence of 45 s and a pixel size of 0\arcsec.5.

The NVST and SDO data were processed using their respective standard data reduction pipelines. To co-align the datasets, the NVST data were cross-correlated with the SDO/AIA data. This alignment process involved two steps: first, the roll angle was corrected by matching the orientation of an observed jet; second, the heliocentric coordinates were aligned by matching the cusp structure visible in both the NVST H$\alpha$ and SDO/AIA images.

Data from six AIA EUV channels (94~\AA, 131~\AA, 171~\AA, 193~\AA, 211~\AA, and 335~\AA) were employed to construct the differential emission measure (DEM) and diagnose the plasma's thermal properties. The average electron number density, $n_e$, in the reconnection region is estimated using the relation $n_e = \sqrt{\mathrm{EM}/H}$, assuming a filling factor of unity. Here, EM is the emission measure derived from the DEM analysis, and $H$ is the plasma depth along the line of sight (LOS). For this study, the LOS depth $H$ is assumed to be the height of the low corona above the photosphere, estimated as $\sim$3~Mm.

It should be noted that the atmospheric seeing for this NVST H$\alpha$ dataset was unsteady, which adversely affected observations during the pre- and post-reconnection stages. However, the data still clearly captured the evolution of the dissipating current sheet. Therefore, in this study, we primarily utilize the NVST H$\alpha$ data to analyze the development of this current sheet.

\section{Results} 
This study focuses on a small-scale jet event on November 24, 2020, in a quiet-Sun region near heliocentric coordinates ($-730\arcsec$, $-430\arcsec$). Figure~1 presents the temporal evolution of the region of interest (ROI) in the layered solar atmosphere, within a field of view (FOV) of $23\arcsec \times 35\arcsec$. From top to bottom, the panels display observations from the SDO/AIA 94~\AA, SDO/AIA 171~\AA, and NVST H$\alpha$ channels. The 94~\AA\ and 171~\AA\ channels are sensitive to plasma at temperatures of $10^{6.8}$~K and $10^{5.8}$~K, respectively, while the H$\alpha$ line captures cooler chromospheric material \citep{2012ApJ...749..136L}. This combination of multi-wavelength observations enables a comprehensive diagnosis of the dynamics of plasma ejections at various temperatures and the evolution of magnetic fields. The time series images from left to right in Figure~1 display the whole development process of the small-scale jet. The first to fifth columns correspond to the initial stage, H$\alpha$ jet forming, EUV jet forming, blob ejection, and blob propagation. The LOS magnetic field map is overlaid on Figure~1a to show the magnetic polarity distribution at the initial stage. It reveals that the source region was located in a mixed-polarity zone. Three key magnetic polarities related to this event are labeled P1, P2, and N (see Figure~1a). There existed smaller negative magnetic fragments between P1 and N, including pre-existing and newly split ones (see Figure3.mp4), whose motions are highly related to the development of the current sheet in association with the observed small-scale jet. For convenience, we thereafter refer to these smaller negative magnetic fragments as N1.
 
 \begin{figure*}
 	\epsscale{1}
 	\figurenum{1}
 	\plotone{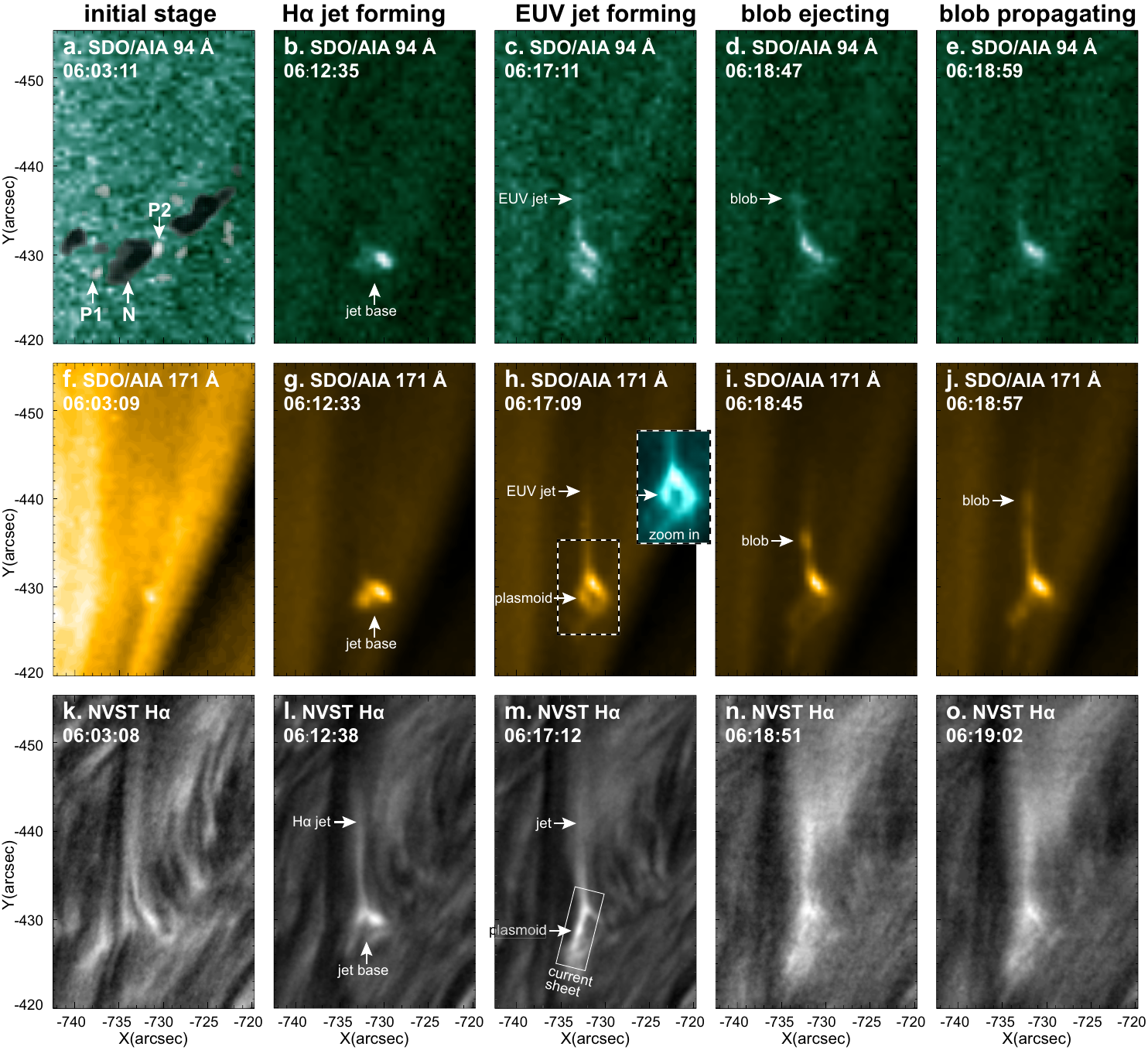}
 	\caption{Multi-wavelength evolution of the small-scale reconnection jet. From top: SDO/AIA 94 \AA, SDO/AIA 171 \AA\ and NVST H$\alpha$ observations. Time runs left to right. The distributions of magnetic polarities at 06:12 UT are overlaid on (a), with P1, P2, and N1 marked; The positive and negative magnetic fields are saturated at 100 and -150 G, respectively. Key observational features, such as jet base, cool jet, hot jet, plasmoid, current sheet, and blob, are annotated in corresponding panels. The subgraph in (h) shows SDO/AIA 131 \AA\ image. The white box in (m) represents the FOV of Figure 2. An animation spanning 06:00--06:30 UT is available online, which doesn't include overlaid SDO/HMI observations and annotations of key observational features. 
 		\label{fig1}}
 \end{figure*}
 
Initially, no prominent atmospheric features are visible in the source region (Figure~1a). The event commences at approximately 06:12:35~UT with the formation of a small-scale bright jet base, about 4$\arcsec$ in length, which appears as a bright patch with its right side being more intense than the left (Figure~1b). About five minutes later, a thin, collimated jet is ejected from this bright patch, forming a classic inverted-Y structure (Figure~1c). The appearance of this morphology strongly indicates that the jet was driven by magnetic reconnection, a mechanism well-established in numerous simulation and observational studies \citep[e.g.,][]{2007Sci...318.1591S, 2011ApJ...731...43N, 2011PhPl...18k1210S, 2011ApJ...738L..20H, 2011ApJ...735L..43S, 2019ApJ...885L..11S, 2021RSPSA.47700217S, 2012ApJ...745..164S, 2014ApJ...786..151S, 2014ApJ...795..130S, 2018ApJ...854...92T, 2023ApJ...945...96Y, 2023ApJ...942...86Y, 2024ApJ...964....7Y, 2023ApJ...942L..22D, 2024ApJ...968..110D, 2024ApJ...962L..35S, SHARMA2025,2021ApJ...912L..15T,2024ApJ...970...77L,2021ApJ...911...33C}. Subsequently, between 06:17:11~UT and 06:18:59~UT, a plasma blob of approximately 2~Mm$^2$ appears on the jet spire (Figure~1d). This blob propagates upward with the spire and is noticeably hotter than its surroundings, as confirmed by the DEM analysis presented in Figure~4. It rapidly dissipates within 24 seconds (Figure~1e), a lifetime significantly shorter than that of typical, larger plasma blobs \citep{2014A&A...567A..11Z, 2017ApJ...841...27N, 2017ApJ...851...67S}.

\begin{figure*}[t]
	\epsscale{0.8}
	\figurenum{2}
	\plotone{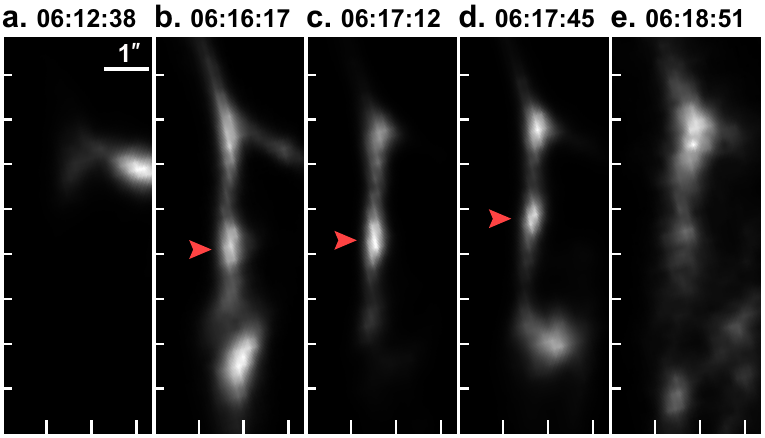}
	\caption{Evolutionary images of the development of the reconnection current sheet. Panels (a-e) are zoomed-in versions of NVST H$\alpha$ observations, whose FOV is represented by the white box in Figure 1m. Their contrasts are adjusted to better visualize the development of the reconnection current sheet. Red arrows in panels (b-d) symbolize the propagating process of a plasmoid.
		\label{fig2}}
\end{figure*}

A comparison between the first and second rows of Figure~1 reveals that the jet's temporal evolution in the SDO/AIA 171~\AA\ channel is broadly similar to that in the 94~\AA\ channel, with both showing the jet's formation and the blob's ejection. However, a closer inspection reveals two subtle differences. First, during the jet's formation, the 171~\AA\ image displays a brightening across the entire jet base in a distinct triangular shape (Figure~1g), which contrasts with the more partial brightening seen in the 94~\AA\ image (Figure~1b). This morphological difference suggests temperature stratification within the jet base. Second, a blob-like feature is already present at the jet base in the 171~\AA\ image (Figure~1h) before the ejection of the plasma blob along the spire (seen in Figure~1d). This pre-existing feature is later identified as a plasmoid.

Since the NVST H$\alpha$ channel probes deeper and cooler layers of the solar atmosphere than the EUV channels, it is used here to diagnose the jet's origin, with the results shown in the third row of Figure~1. Initially, the H$\alpha$ image reveals several cool magnetic threads in the source region (Figure~1k). In the subsequent stage, the H$\alpha$ image shows both the triangular-shaped jet base, similar to that in the EUV wavelengths, and a faint, thin jet spire. This spire is not yet visible in the corresponding EUV images, a difference made clear by comparing Figure~1l with Figures~1b and 1g. These observations suggest that magnetic reconnection and the initial plasma ejection had already commenced, but the ejected plasma was still too cool to be detected in the hot EUV channels.

During the EUV jet formation stage, the NVST H$\alpha$ images resolve the fine structure of the jet base more clearly. An elongated, sheet-like structure is evident within the jet base (Figure~1m), which, when compared with the LOS magnetogram (Figure~1a), is confirmed to lie above the polarity inversion line (PIL) between the magnetic polarities P1 and N. The blob identified in the AIA 171~\AA\ and 131~\AA\ images (Figure~1h) is also resolved in the H$\alpha$ image (Figure~1m). We observe this blob moving upward along the bright, sheet-like structure; however, a corresponding plasma blob is not detected along the jet spire in the H$\alpha$ images.

To better visualize the development of the elongated sheet-like structure within the jet base, we present a sequence of zoomed-in H$\alpha$ images in Figure~2. The sequence begins at 06:12:38~UT, during the initial formation of the H$\alpha$ jet, when the feature is still short and not yet clearly sheet-like (Figure~2a). The structure then rapidly extends between 06:16:17~UT and 06:17:45~UT, evolving into the distinct elongated sheet (see Figures~2b--d, Figure~1, and the online animation). Coinciding with this rapid elongation, a bright plasma blob with dimensions of about $1\arcsec \times 1\arcsec$ appears near the middle of the structure. This blob is observed moving upward along the sheet, traveling a projected distance of approximately 0.58~Mm in 88~seconds (from 06:16:17~UT to 06:17:45~UT, indicated by red arrows in Figures~2b--d). This corresponds to an average projected speed of about 6.6~km~s$^{-1}$. The plasma blob subsequently faded and disappeared by 06:18:51 UT (Figure~2e).

\begin{figure*}
	\epsscale{0.9}
	\figurenum{3}
	\plotone{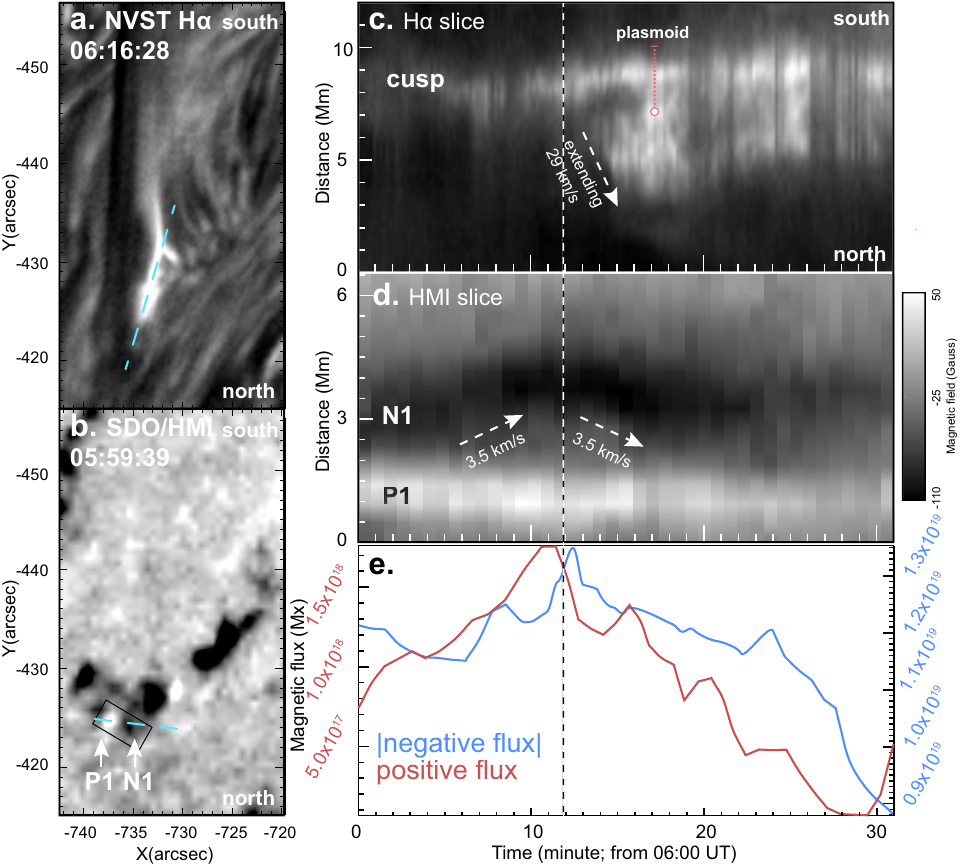}
	\caption{Analysis of the development of reconnection current sheet. Panels (a-b) show the slice paths and flux integrated region. Time-distance diagrams from H$\alpha$ (c), HMI (d), and simultaneous magnetic flux profiles (e) reveal two distinguishable phases of developmental reconnection current sheet, in which the vertical dotted line represents the phase shift time of current sheet length as well as magnetic fluxes ($\sim$06:10 UT). The color bar in (d) also applies to panel (b). An animation spanning 06:00--06:30 UT is available online, which only includes SDO/HMI observations and its slice.
		\label{fig3}}
\end{figure*}

\begin{figure*}
	\epsscale{0.9}
	\figurenum{4}
	\plotone{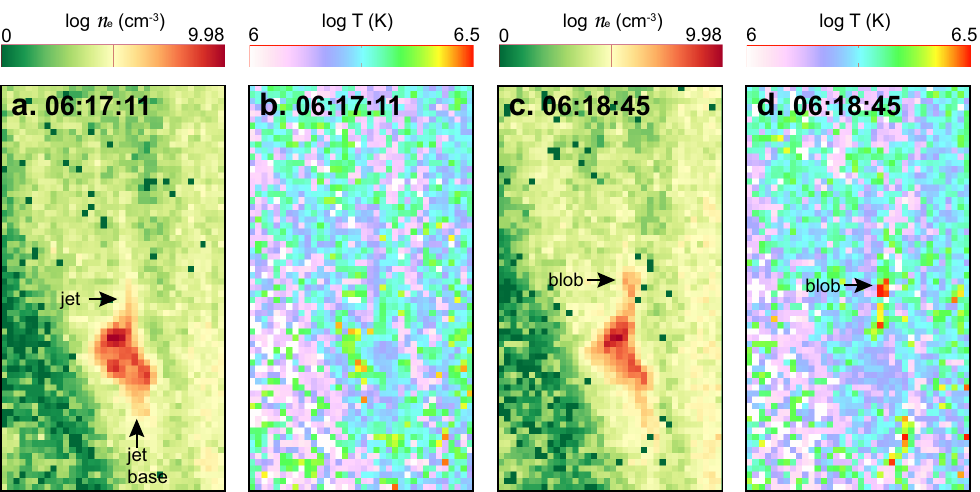}
	\caption{DEM analysis before and after the blob ejection. Panels (a and c) and (b and d) represent density and averaged temperature maps, respectively. Key observational features, such as jet, jet base, and blob, are annotated in the corresponding panels. 
		\label{fig4}}
\end{figure*}

To quantify the temporal development of the sheet-like structure, we constructed a time-distance map from the NVST H$\alpha$ images along the slice shown in Figure~3a. The resulting map (Figure~3c) reveals that the structure's south end, initially at $y \approx 8$~Mm, gradually shifts to $y \approx 10$~Mm. The evolution can be divided into two distinct phases. In the first phase (06:07~UT to 06:12~UT), the reconnection dissipation region manifests as a short, faint brightening within the jet base, accompanied by an H$\alpha$ cool jet. In the second phase (06:12~UT to 06:15~UT), the reconnection dissipation region extended rapidly by about 6~Mm at an average speed of approximately 29~km~s$^{-1}$ to form the long sheet-like structure. This phase was followed by a sequence of events: a plasmoid (Figure~1h and m), an EUV hot jet (Figure~1c and h), and an ejected plasma blob (Figure~1d, i and j) were detected sequentially.

To investigate the photospheric driving mechanism, we analyzed the evolution of the current sheet's footpoints (N1 and P1) using a time-distance map constructed from SDO/HMI magnetograms (Figure~3d). The map, generated along the slice in Figure~3b, shows the polarity inversion line (PIL) situated at $y \approx 2$~Mm. While the positive polarity P1 remains stationary at $y \approx 1$~Mm, the negative polarity N1, initially at $y \approx 3$~Mm, first moved away from P1 between 06:05~UT and 06:12~UT at a speed of $\sim$3.5~km~s$^{-1}$ (hereafter, the separating phase). Subsequently, N1 reversed direction and moved toward P1 at a similar speed (hereafter, the converging phase). A quantitative analysis of the magnetic flux within the boxed region in Figure~3b corroborates this behavior. As plotted in Figure~3e, the positive (red) and negative (blue) magnetic fluxes both show an initial increasing trend, peaking around 06:12~UT, after which they begin to decrease, indicating flux cancellation. This pattern of flux variation is consistent with the separating and subsequent converging motions of the magnetic polarities seen in Figure~3d.

Synthesizing these results reveals a strong correlation between the photospheric magnetic activity and the chromospheric and coronal phenomena. The rapid extension of the sheet-like structure corresponds directly to the converging motion of the opposite magnetic polarities (Figure~3d) and the onset of the flux cancellation phase (Figure~3e). A clear distinction emerges between the two phases: the separating phase is associated with the appearance of the triangular bright structure and the H$\alpha$ cool jet. In contrast, the converging and flux cancellation phase is associated with the formation of the elongated sheet carrying a plasma blob, the ejection of the hot EUV jet, and the appearance of the hot plasma blob along the jet's spire.

We employed the DEM method to diagnose the plasma properties before ($\sim$06:17:11~UT) and after ($\sim$06:18:45~UT) the appearance of the upward-propagating plasma blob. The results are displayed in Figure~4, which presents electron density maps (panels a and c) and emission-measure-weighted temperature maps (panels b and d) for these two moments. Figure~4a and 4b correspond to the time before the blob's appearance, while Figure~4c and 4d correspond to the time after. At 06:17:11~UT, before the blob's appearance, the density map clearly resolves the jet and its base in a typical inverted-Y morphology (Figure~4a). The electron density is highest at the top of the triangular cusp ($n_e \sim 10^{10}$~cm$^{-3}$) and decreases significantly up into the jet spire ($n_e \sim 10^5$~cm$^{-3}$). In contrast, the jet's base and spire can not be discernible in the corresponding temperature map (Figure~4b). By 06:18:45 UT, the density map reveals the upward-propagating plasma blob clearly as indicated by the arrow in Figure~4c. Its density ($n_e \sim 10^8$~cm$^{-3}$) is comparable to that of the surrounding jet plasma, showing no significant contrast. Instead, it is the feature's distinct blob-like morphology that allows it to be identified. The blob is, however, prominent in the temperature map (Figure~4d). It appears as a localized, high-temperature structure with an average temperature of $\sim 10^{6.5}$~K. This distinct thermal signature confirms that the observed feature is a hot plasma structure.

\section{Interpretation}
Based on the preceding analysis, we interpret the elongated sheet-like structure as the reconnection current sheet responsible for the jet's generation. The formation and evolution of this current sheet are intimately linked to the dynamics of the photospheric magnetic polarities, with distinct atmospheric responses corresponding to the increase and subsequent cancellation phases of the photospheric magnetic flux.

We propose the following physical model to explain the generation and evolution of the observed jet, in which the motions of the photospheric magnetic polarities govern the key processes.  During the separating phase, the motion of N1 and P1 away from each other likely reduced the magnetic stress in the overlying atmosphere, limiting the development of a large-scale current sheet. Consequently, reconnection was confined to a mild phase, producing a short, faint brightening region (Figure~3c) and a relatively cool H$\alpha$ jet (Figure~1l), despite the overall increase in magnetic flux. Conversely, during the converging phase, the approach of N1 toward P1 enhanced the magnetic stress, facilitating the rapid extension of the dissipation region into a thin, elongated current sheet. This elongation triggered a tearing-mode instability, leading to the formation of a magnetic island (a plasmoid) and initiating an intense phase of fast magnetic reconnection, which generated the hotter EUV jet. Once the magnetic reconnection intensified, the magnetic flux from the inflow region was efficiently converted to the outflow. The resulting enhancement of magnetic pressure in the downflow region would drive the submergence of newly formed, closed loops, manifesting as the significant magnetic flux cancellation observed in the HMI magnetograms \citep{1985SoPh..100..397Z,1987ARA&A..25...83Z,1999ApJ...515..435L}. The total canceled flux $\phi$ between 06:10~UT and 06:30~UT was approximately $10^{18}$~Mx, implying a cancellation rate of $\sim 10^{15}$~Mx~s$^{-1}$. From this, we estimate a lower limit for the released magnetic energy during this intense phase as $E = (B \phi L) / (8\pi) \approx 10^{26}$~erg, corresponding to an energy release rate of $\sim 10^{23}$~erg~s$^{-1}$. Here, $L \approx 6\arcsec$ is the current sheet length and $B \approx 50$~G is the magnetic field strength, estimated from the photospheric magnetic fields involved in the cancellation (Figure~3d).

\begin{figure*}
	\epsscale{1}
	\figurenum{5}
	\plotone{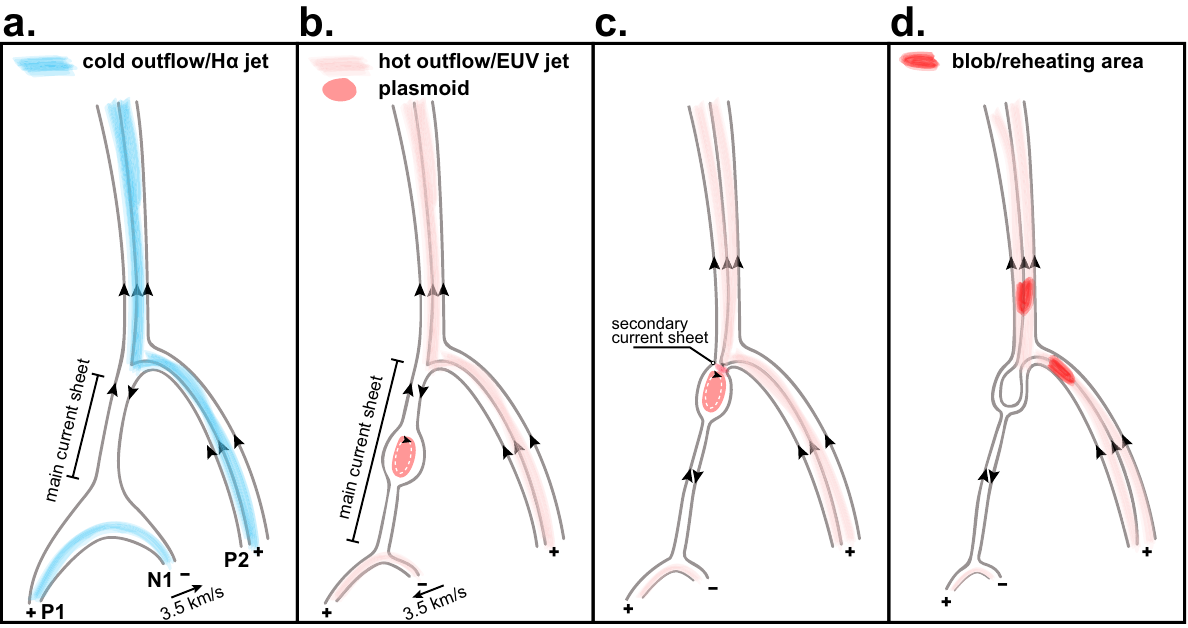}
	\caption{A cartoon interpretation of the present event. (a) shows the magnetic skeleton at the cool jet stage, in which blue areas represent cold (H$\alpha$) reconnection outflow. (b) displays the lengthening of the reconnection current sheet, and correspondingly formed plasmoid, where red areas represent hot (EUV) reconnection outflow. (c) illustrates the upward propagating process of the plasmoid as well as its collision with the upper cusp and the in-between secondary current sheet. (d) represents the merging process of the plasmoid with the upper cusp and resulting blob. Note that this cartoon incorporates certain features not directly observed in our observation.
		\label{fig5}}
\end{figure*}

The DEM analysis at 06:18:45 UT reveals that the plasma blob in the jet spire is significantly hotter than the ambient jet plasma. This suggests that the heating mechanisms for the jet and the blob are different. We propose that the blob's temperature can be conceptually decomposed into two parts: (1) a main component, comparable to the temperature of the surrounding jet and caused by the primary magnetic reconnection in the main current sheet, and (2) a secondary component, representing the excess temperature that requires an additional heating process. Previous studies suggest a viable mechanism for such additional heating: the collision of a plasmoid with the overlying cusp structure \citep{2008A&A...477..649B, 2010ApJ...713.1292M, 2011ApJ...733..107K, 2012A&A...541A..86K, 2017ApJ...841...27N}. A collision between the anti-parallel magnetic fields of the plasmoid and the cusp can form a secondary current sheet. Reconnection within this secondary sheet would fragment the plasmoid's O-shaped field lines, facilitate its merger with the cusp field, and generate localized, enhanced heating. We therefore infer that the upward-propagating plasmoid (the blob within the main current sheet) interacts with the outflow region, forming such a secondary sheet whose energy release provides the additional heating needed to explain the secondary temperature component of the hot blob in the spire.

To illustrate this comprehensive model, we present a schematic cartoon in Figure~5. The cartoon depicts key representative features, including the three magnetic polarities (P1, P2, and N1), the motion of N1, the current sheet, the upward-propagating plasmoid within it, the jet, and the final hot plasma blob in the spire. The schematic also incorporates inferred features not directly resolved in our observations, such as post-reconnected loops, an anti-sunward propagating blob, and the aforementioned secondary current sheet. The post-reconnected loops may have been obscured by overlying chromospheric fibrils, while the secondary current sheet was likely too small to be resolved by AIA. Including these elements provides a more complete physical picture of the event.

The model begins with the formation of the cool H$\alpha$ jet during the mild reconnection stage (Figure~5a). At this point, the current sheet forms between interacting magnetic loops, but its aspect ratio is insufficient to trigger fast, tearing-mode reconnection. The resultant mild energy release produces the cool jet and the triangular brightening. Subsequently, the converging motion of N1 (Figure~3d and Figure~5a) squeezes the magnetic field, forcing reconnection between the field lines rooted in P1 and N1 and forming a longer, thinner current sheet (Figure~5b). Once the sheet's aspect ratio crosses the critical threshold for tearing instability, a plasmoid forms, igniting a fast reconnection regime and generating the hot EUV jet \citep{1963PhFl....6..459F, 2001EP&S...53..473S, 2009PhPl...16k2102B, 2012ApJ...745L...6T}. Newly-formed post-reconnected loops then submerge, appearing as magnetic cancellation (Figure~3e). This sequence is consistent with our observation that the plasmoid, the hot EUV jet, the converging motion of N1, and magnetic cancellation all appeared concurrently after the current sheet's significant elongation.

Following its formation, the plasmoid moves upward along the current sheet and collides with the cusp structure above (Figure~5c). Because their magnetic fields are antiparallel, a secondary current sheet forms between them. Reconnection here releases additional energy, resulting in localized heating and potential re-acceleration of plasma already preheated by the main current sheet \citep{2008A&A...477..649B, 2010ApJ...713.1292M, 2011ApJ...733..107K, 2012A&A...541A..86K}. This "double heating" mechanism can account for the observed hot plasma blob in the jet spire. A key prediction of this model is that the appearance of the hot blob in the spire should follow the disappearance of the plasmoid in the main current sheet. Upon re-examining Figure 1 and the accompanying online video, we confirm a clear temporal correlation between these two events. This finding provides compelling support for our physical interpretation.

\section{Conclusions and Discussions}
Using high-resolution, multi-wavelength observations from the NVST and the SDO, we presented a detailed analysis of a small-scale chromospheric jet driven by plasmoid-mediated magnetic reconnection in the chromosphere. Our results reveal that the reconnection was driven by the motion of photospheric magnetic polarities and proceeded in two distinct stages: a mild phase followed by a more intense one. The mild reconnection phase coincided with the initial separating motion of the photospheric footpoints, which led to the formation of a short current sheet, a small triangular brightening, and a cool H$\alpha$ jet. This stage, characterized by moderate energy release, may represent the buildup process of the reconnection dissipation region. The intense reconnection phase commenced as the footpoint N1 began to converge toward P1, a transition marked by the current sheet's rapid elongation, the appearance of a plasmoid within it, and the ejection of a hot EUV jet. These phenomena indicate that the plasmoid, formed via a tearing instability as the current sheet's aspect ratio decreased, mediated a transition to a fast reconnection regime.

We interpret the observed magnetic flux cancellation as evidence for the submergence of newly formed, closed reconnection magnetic loops in the lower outflow region, a process that began concurrently with the onset of fast reconnection. Furthermore, we propose that the high-temperature plasma blob observed in the jet spire results from secondary heating, caused by reconnection between the upward-moving plasmoid and the overlying magnetic cusp. These observational results provide new insights into the development of small-scale solar jets. Our cartoon model, which incorporates both flux cancellation and a two-step reconnection process, naturally explains all of these observed characteristics.

Our observations highlight the morphological transition of the current sheet from a short to an elongated structure, a change that was temporally correlated with the initial separating and later converging motions of the current sheet's photospheric footpoints. The initial separating motion corresponded to the onset of reconnection. As the photospheric footpoints entered the converging phase, increased magnetic pressure caused the current sheet's width-to-length ratio to decrease significantly. When this ratio reached a critical threshold, it triggered the onset of the tearing instability, dramatically enhancing the magnetic energy release. These findings underscore the critical role of photospheric footpoint motion in controlling current sheet formation, triggering plasma instabilities, and driving eruptive activity in the solar atmosphere.

The substantial increase in the energy release rate is observationally evidenced by the transition from a cool H$\alpha$ jet to a hot EUV jet and by the switch in magnetic flux evolution from increase to cancellation. The flux cancellation itself suggests stronger reconnection outflows resulting from the intensified energy release. This energy release rate appears intimately linked to the current sheet's length. During the mild phase, the short current sheet produced only a cool jet and a minor brightening. In contrast, during the intense phase, the significantly elongated current sheet and the onset of tearing instability resulted in a hot EUV jet and a larger, brighter feature. The total released magnetic free energy is estimated to be on the order of $10^{26}$~erg (with a peak rate of $\sim10^{23}$~erg~s$^{-1}$), which is sufficient to offset radiative losses over an area of $\sim$1~Mm$^2$ \citep[assuming a loss rate of $\sim10^6$~erg~cm$^{-2}$~s$^{-1}$;][]{1977ARA&A..15..363W}.

The intensification of the reconnection heat can be attributed to two complementary factors. First, the tearing instability associated with the current sheet's lengthening is crucial. The resulting plasmoids are known to accelerate the reconnection process and can initiate secondary energy release episodes \citep{2001EP&S...53..473S, 2008A&A...477..649B, 2009PhPl...16k2102B, 2010ApJ...713.1292M, 2011ApJ...733..107K, 2012A&A...541A..86K}. Second, the energy release rate is also expected to scale with the physical length of the current sheet itself \citep[e.g.,][]{2024MNRAS.534.3133P}. Therefore, the rapid elongation of the current sheet played a direct and additional role in intensifying the energy release.

Finally, the propagation speed of the observed plasmoid warrants discussion. The measured speed of $\sim$6.6~km~s$^{-1}$ is notably slower than the tens of km~s$^{-1}$ reported in other chromospheric reconnection events \citep[e.g.,][]{2012ApJ...759...33S, 2021A&A...646A..88N, 2023A&A...673A..11R}. This discrepancy highlights that a plasmoid's speed does not necessarily approach the local Alfv\'en speed and can vary significantly between events. The speed is ultimately modulated by the net magnetic tension force, which arises from the difference in reconnection rates at the plasmoid's leading and trailing edges. The local Alfv\'en speed merely sets the upper theoretical limit for this propagation \citep{2008A&A...477..649B}.

\begin{acknowledgments}
The authors wish to thank the anonymous referee for their insightful suggestions, which have significantly strengthened this paper. We gratefully acknowledge the science teams of the Solar Dynamics Observatory (SDO) and the New Vacuum Solar Telescope (NVST) for their open data policy and the high-quality observations used herein. This research was supported by the Natural Science Foundation of China (under grants 12173083 and 12163004), the Shenzhen Key Laboratory Launching Project (No. ZDSYS20210702140800001), and the Specialized Research Fund for the State Key Laboratory of Solar Activity and Space Weather.
\end{acknowledgments}

\bibliographystyle{aasjournal}
\bibliography{bibfile}

\end{document}